\documentclass{article}
\usepackage[utf8]{inputenc}
\usepackage{authblk}
\usepackage{setspace}
\usepackage[margin=1.25in]{geometry}
\usepackage{graphicx}
\graphicspath{ {./figures/} }
\usepackage{subcaption}
\usepackage{amsmath}

\usepackage[
    colorlinks=true,
    linkcolor=black, 
    citecolor=black, 
    urlcolor=blue    
]{hyperref}

\usepackage[style=nejm, 
citestyle=numeric-comp,
sorting=none]{biblatex}
\title{Phase-Stable Self-Modulation for GHz Continuous-Wave Ultrafast X-Ray Free-Electron Lasers}

\author[1]{Junhao Liu}
\author[1,2,*]{Zhen Wang}
\author[1]{Lanpeng Ni}
\author[3]{Yujie Lu}
\author[1,2,*]{Chao Feng}
\author[1,2]{Zhentang Zhao}

\affil[1]{Shanghai Advanced Research Institute, Chinese Academy of Sciences, Shanghai 201210, China}
\affil[2]{University of Chinese Academy of Sciences, Beijing 100049, China}
\affil[3]{Zhangjiang Laboratory, Shanghai 201204, China}
\affil[*]{Address correspondence to: \href{mailto:wangz@sari.ac.cn}{wangz@sari.ac.cn} (Z.W.); \href{mailto:fengc@sari.ac.cn}{fengc@sari.ac.cn} (C.F.)}

\date{}

\begin{document}

\maketitle

\begin{abstract}
High-brightness femtosecond-to-attosecond pulses are indispensable for probing electron dynamics on their fundamental temporal scales. X-ray free-electron lasers (XFELs) at high repetition rates will facilitate high-statistics measurements and time-resolved studies that were previously inaccessible. Although energy recovery linacs (ERLs) are well suited for high-repetition-rate operation, their relatively low peak current poses a major challenge for generating intense ultrashort X-ray pulses. Here, we propose a completely laser-free scheme that fundamentally overcomes this bottleneck through a continuous, phase-stable self-modulation process. By interacting with its own coherently emitted terahertz radiation within a helical wiggler, the electron bunch naturally accumulates a robust, few-cycle energy modulation in its core, even when starting with the intrinsically low peak current typical of ERLs. A downstream dispersion chicane subsequently converts this energy modulation into an isolated, exceptionally sharp current spike. Start-to-end simulations based on a 1~GeV ERL light source demonstrate the feasibility of generating isolated soft X-ray pulses with an average peak power exceeding 4~GW and a pulse duration of about 1~fs at an unprecedented 1.3~GHz repetition rate. The proposed scheme offers a highly practical pathway for advancing ultrafast X-ray generation into the true continuous-wave regime, with transformative implications for the development of next-generation coherent light sources.
\end{abstract}


\section{Introduction}

The evolution of X-ray free-electron laser (XFEL) technology has revolutionized cutting-edge research across multiple disciplines, driven by its unparalleled brightness and short pulse durations \cite{feng2018review,huang2021features}. In particular, femtosecond X-ray pulses provide a powerful tool for tracking ultrafast electronic redistribution \cite{higley2019femtosecond}, structural evolution \cite{minitti2015imaging}, and phase transitions in molecules and condensed-matter systems \cite{cavalleri2001femtosecond}. To probe such ultrafast phenomena with higher precision, substantial efforts have been made to improve the temporal resolution and coherence of FEL radiation \cite{saldin2008coherence,behrens2014few,huang2017generating,macarthur2019phase,duris2020tunable,zhang2020experimental,prat2023coherent,yan2024terawatt}. Meanwhile, many advanced scientific experiments, such as pump-probe spectroscopy \cite{smolentsev2020taking}, coherent diffraction imaging \cite{wu2021three}, and nonlinear X-ray spectroscopy \cite{kayser2019core}, require not only ultrashort pulses but also a stable, high-density photon flux to achieve high signal-to-noise ratios and resolve weak transient signals. To date, facilities such as FLASH \cite{ackermann2007operation} and the European XFEL \cite{altarelli2015european} have achieved MHz-repetition-rate burst-mode operation \cite{vogt2025flash,wang2023physical}. Furthermore, transitioning to MHz-level repetition rates in continuous-wave (CW) mode would enable a more flexible temporal photon structure and higher average brightness, significantly enhancing FEL performance. While facilities like the Linac Coherent Light Source II (LCLS-II) \cite{galayda2018lcls} at SLAC, the Shanghai High-Repetition-Rate XFEL and Extreme Light Facility (SHINE) \cite{zhu2017sclf}, and the Shenzhen Superconducting Soft X-ray FEL (S$^3$FEL) in China \cite{wang2023physical} have the potential to achieve this goal, increasing the repetition rate severely exacerbates beam loading effects, leading to substantially higher radio-frequency (RF) power demands.

Energy recovery linacs (ERLs) \cite{tigner1965possible,merminga2020energy,adolphsen2022development} are widely considered a promising platform for high-repetition-rate light sources due to their unique combination of high repetition rates and improved power efficiency. They are distinguished by their ability to recover the beam energy by decelerating the electron bunch after the radiation process. This recovered energy is then returned to the linac RF system to accelerate subsequent bunches. With an energy recovery efficiency exceeding 99\%, the RF power demand is drastically reduced, providing a substantial advantage for high-repetition-rate operation. However, the average current is constrained by photoinjector performance and beam break-up (BBU) instabilities \cite{hoffstaetter2004beam,setiniyaz2021filling} within the accelerating cavities. Consequently, the typical average beam current achieved in ERLs is on the order of tens of mA \cite{hoffstaetter2017cbeta,neil2006jlab}, significantly lower than the several hundred mA typical of storage ring light sources. Furthermore, the relatively low bunch charge and peak current limit the applicability of ERLs to advanced operating regimes. Maintaining high beam quality in the recirculating arcs also remains a critical challenge \cite{hall2015measurement, tsai2001suppressing}, limiting achievable bunch compression and thereby constraining the peak current required for high-gain FEL operation at short wavelengths. 

These limitations motivate an alternative strategy in which the beam current is enhanced locally rather than through aggressive compression of the entire bunch. Currently, nonlinear bunch compression \cite{huang2017generating,prat2023coherent,yan2024terawatt} and the enhanced self-amplified spontaneous emission (ESASE) scheme \cite{zholents2005method} stand as the premier techniques for generating intense, femtosecond to attosecond X-ray pulses. While nonlinear compression exploits the intrinsic longitudinal phase-space curvature to form a sharp density peak, conventional ESASE utilizes a high-peak-power external optical laser to imprint a strong energy modulation onto the electron beam within a wiggler, which is subsequently converted into a high-current spike via a dispersive chicane. However, the necessity for such an intense external optical laser in the ESASE scheme fundamentally restricts the achievable repetition rate of the facility. To advance ultrafast X-ray generation into the true continuous-wave GHz regime, it is essential to overcome both the repetition-rate limitations of these laser-driven methods and the intrinsic peak-current constraints of ERLs. While early experimental demonstrations of electron beam self-modulation successfully manipulated the longitudinal phase space without an external laser, they fundamentally relied on the coherent radiation emitted by a pre-existing, localized high-current spike at the bunch tail \cite{macarthur2019phase,duris2020tunable,zhang2020experimental}.

Here, we present a straightforward, advanced self-modulation approach that evolves this mechanism to generate ultrafast X-ray pulses at GHz repetition rates in an ERL-based light source. Rather than depending on an isolated initial tail spike or an external modulation laser, the electron bunch continuously interacts with its own coherently emitted terahertz (THz) radiation field as it propagates along the modulation section. Remarkably, the helical wiggler design allows the beam to naturally accumulate a sufficiently strong, few-cycle energy modulation in its core, even when starting with the intrinsically low peak current typical of ERLs . Crucially, a downstream dispersion chicane efficiently converts this robust energy modulation into an isolated, exceptionally sharp current spike. This localized density enhancement directly bypasses the ERL's low peak current bottlenecks while taking full advantage of the ERL's intrinsically low energy spread, ultralow emittance, and ultrashort bunch durations. Start-to-end (S2E) simulation results demonstrate that this scheme fulfills the stringent requirements for high-gain XFEL emission, opening a practical pathway for generating isolated femtosecond soft X-ray pulses at an unprecedented high repetition rate and unlocking new scientific and industrial frontiers.

\section*{Methods}

\begin{figure}[!b]
\centering
\includegraphics[width=0.8\linewidth]{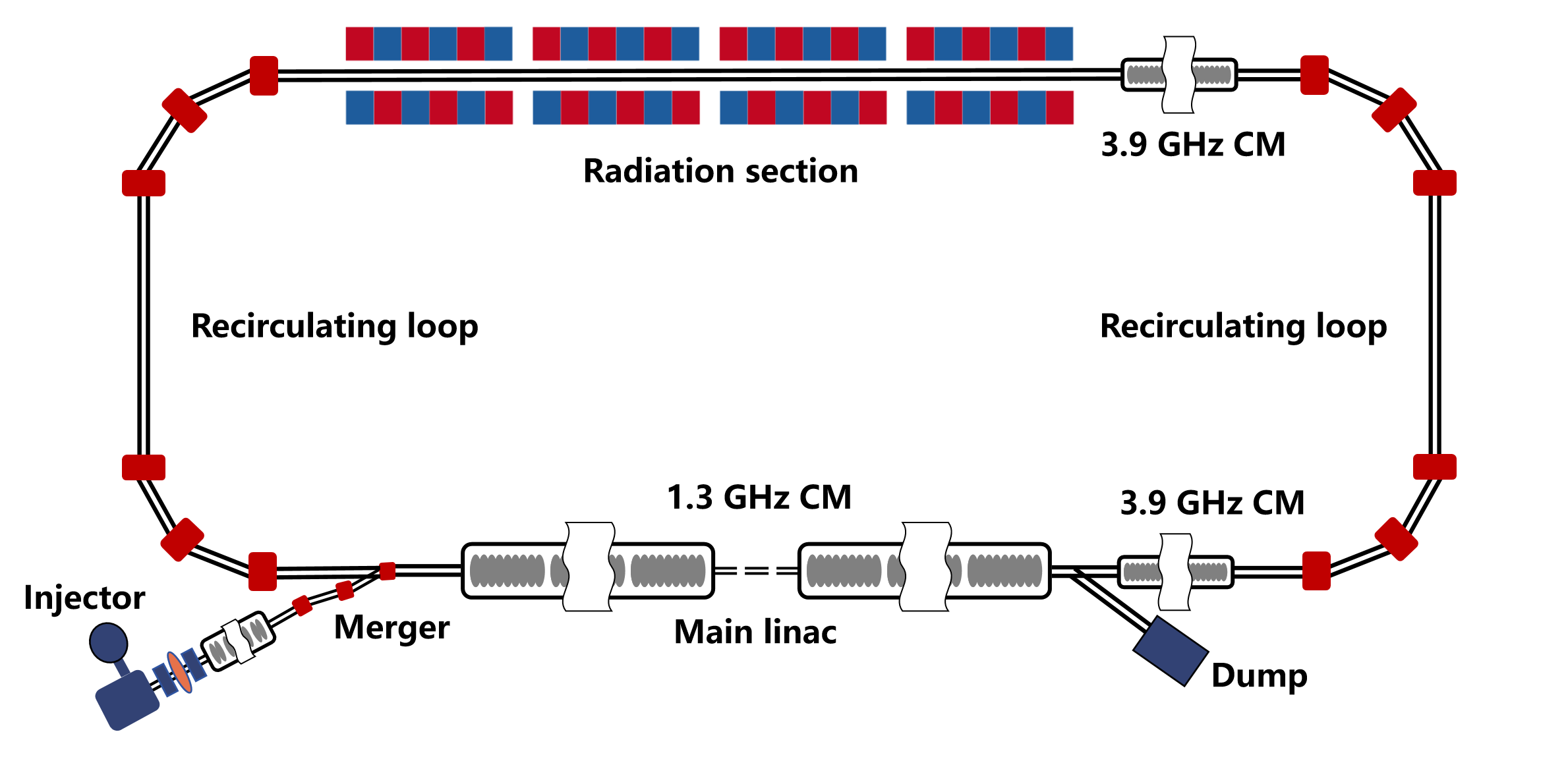}
\caption{The layout of the proposed ERL-based light source.}
\label{fig:erlall}
\end{figure}

\begin{figure}[htbp]
\centering

\begin{subfigure}{.45\textwidth}
  \centering
  \includegraphics[width=\linewidth]{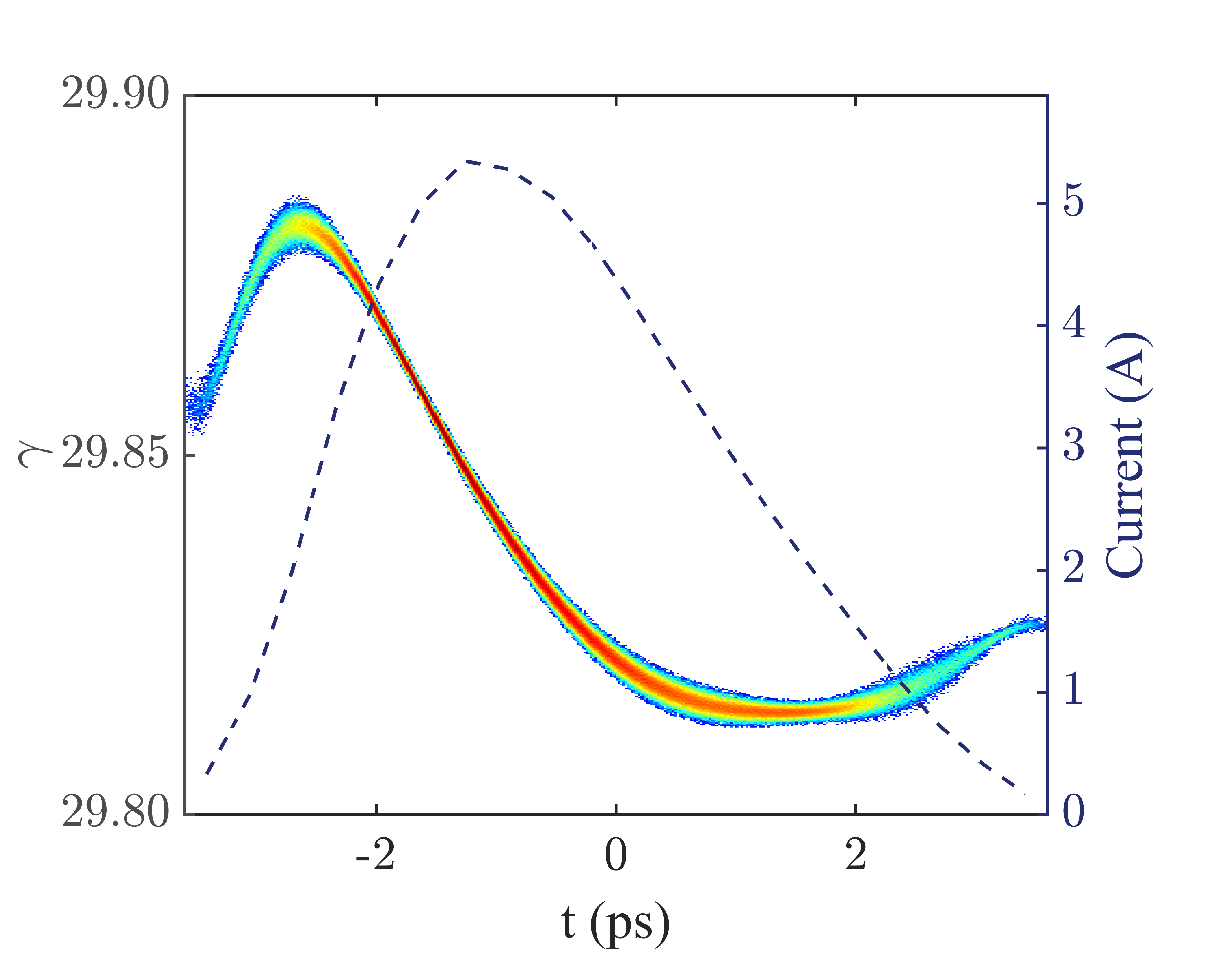}
  \caption{}
  \label{fig:start}
\end{subfigure}
\hspace{0.02\textwidth}
\begin{subfigure}{.45\textwidth}
  \centering
  \includegraphics[width=\linewidth]{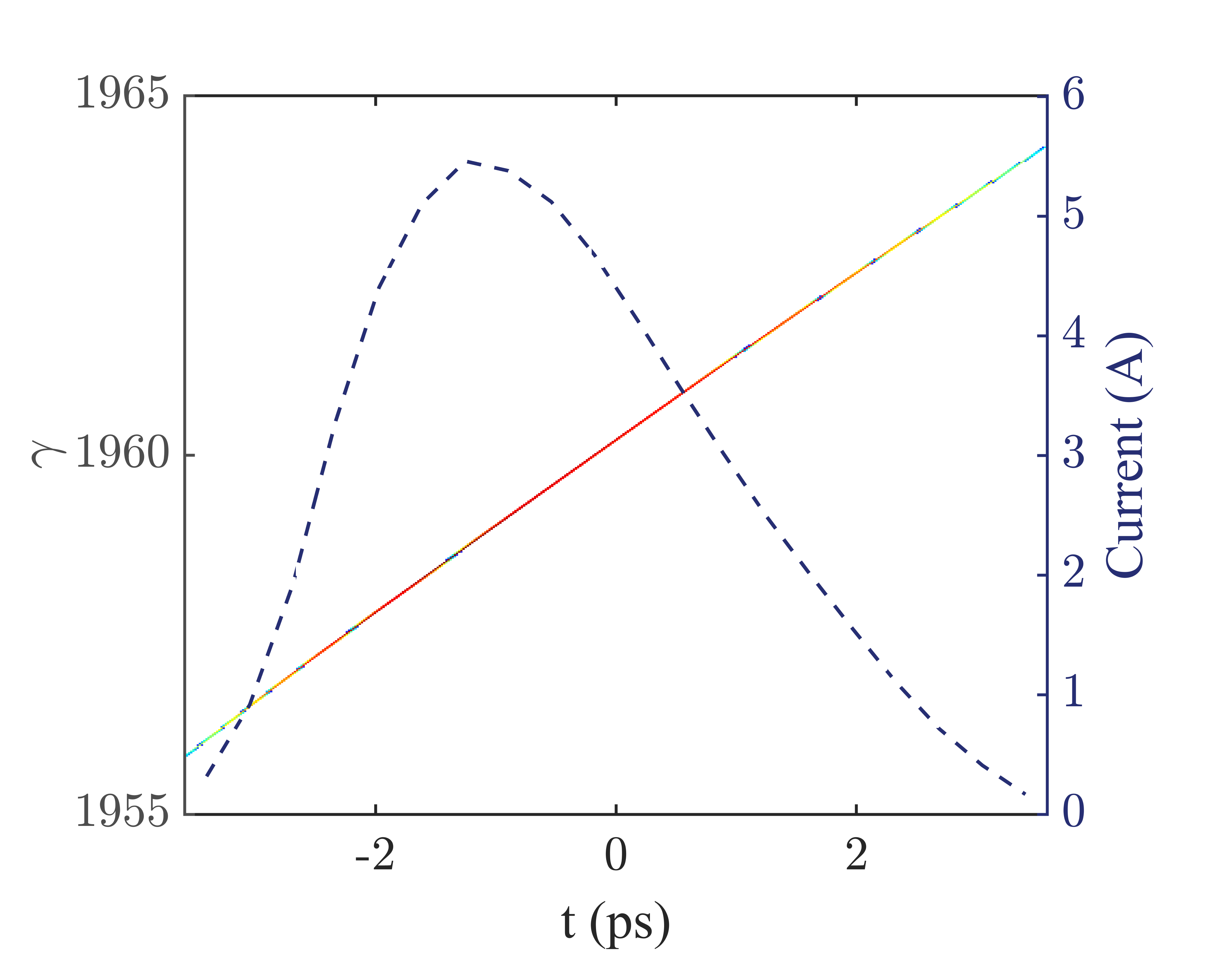}
  \caption{}
  \label{fig:acc}
\end{subfigure}

\vspace{0.2em}

\begin{subfigure}{.45\textwidth}
  \centering
  \includegraphics[width=\linewidth]{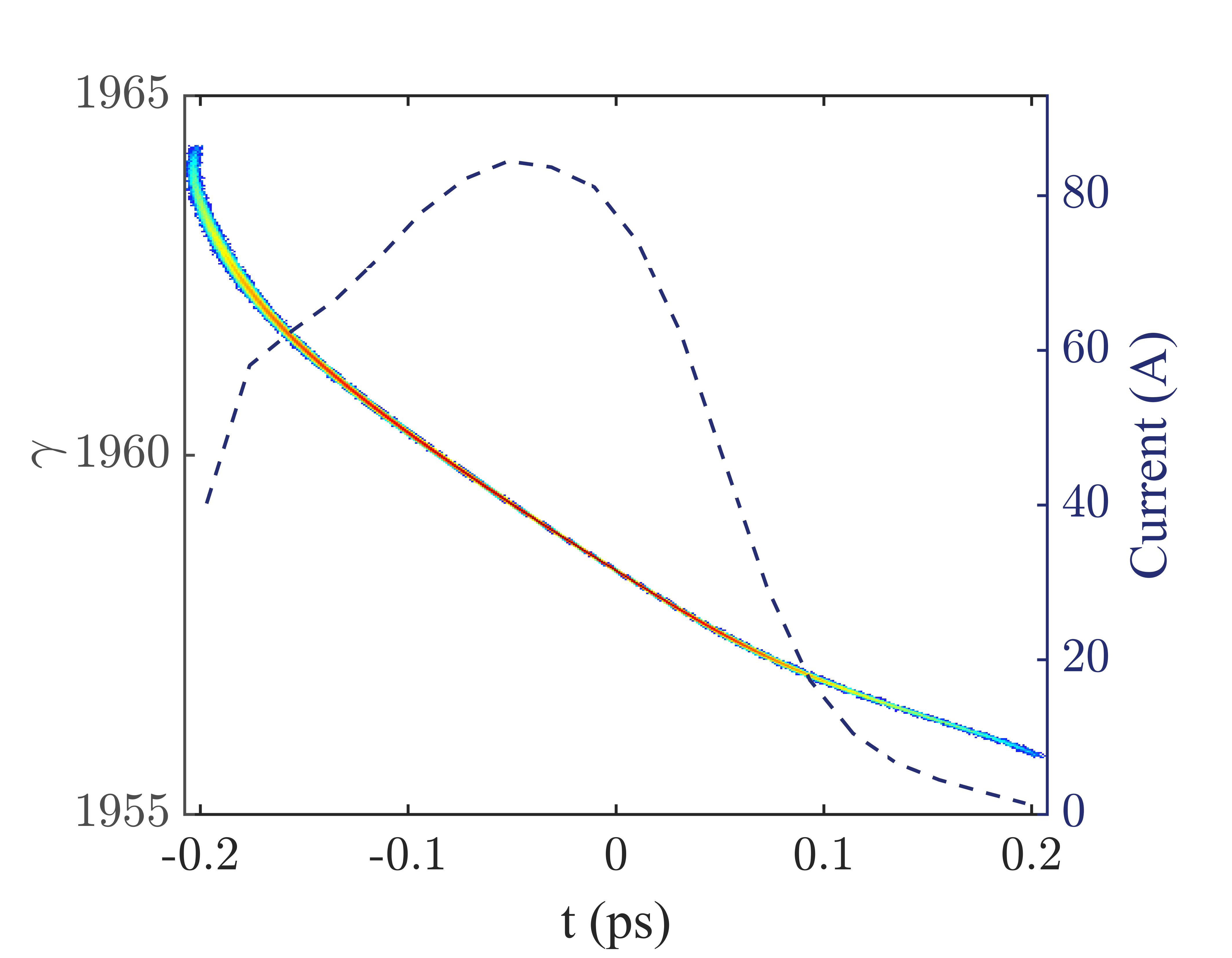}
  \caption{}
  \label{fig:arc}
\end{subfigure}
\hspace{0.02\textwidth}
\begin{subfigure}{.45\textwidth}
  \centering
  \includegraphics[width=\linewidth]{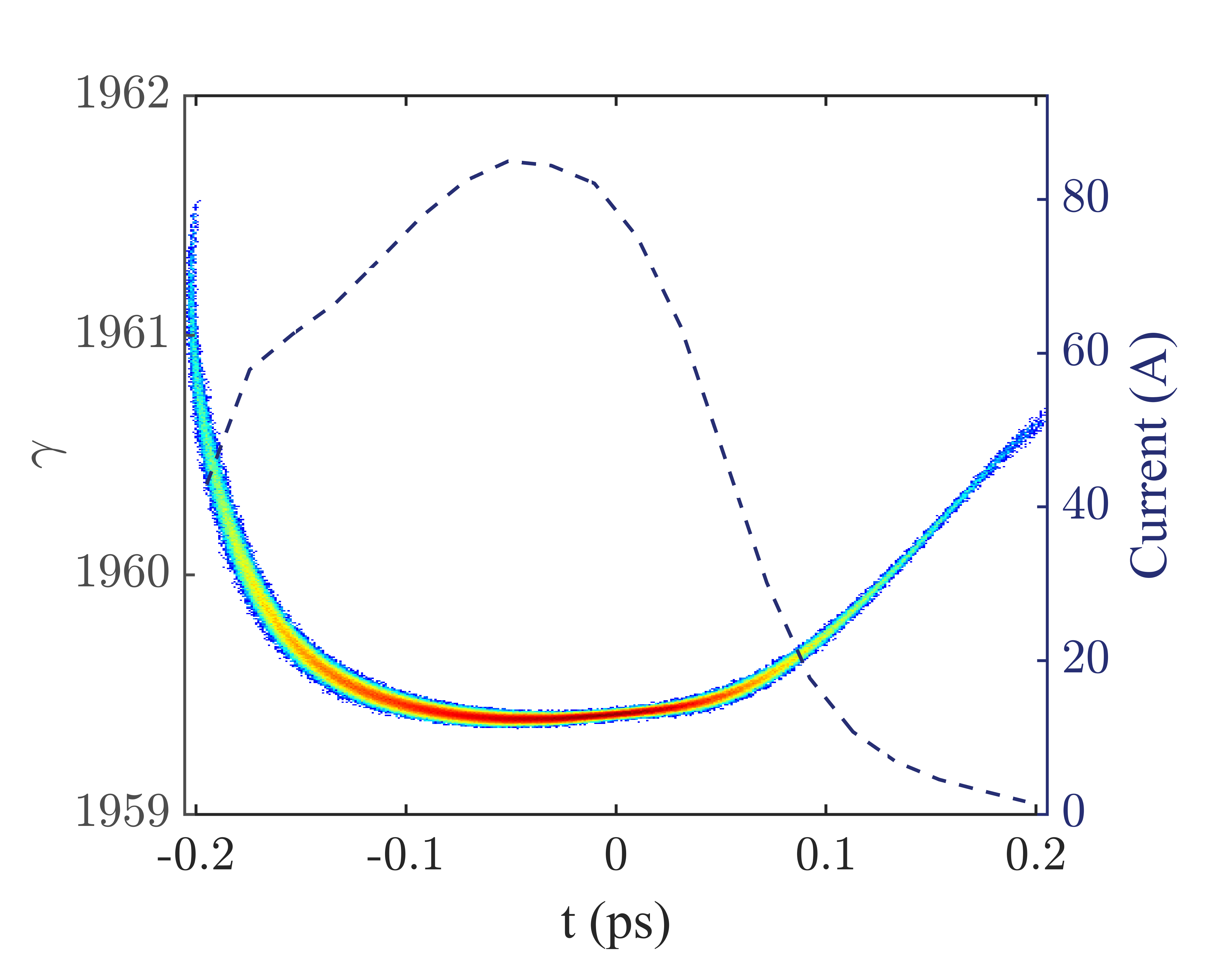}
  \caption{}
  \label{fig:flat}
\end{subfigure}

\caption{Evolution of the longitudinal phase space and current distribution at four locations: (a) after the injector, (b) after the main linac, (c) after the first arc, and (d) at the entrance of the modulator.}
\label{fig:acc_arc}
\end{figure}

Fig \ref{fig:erlall} illustrates the layout of the proposed light source, adapted from a previously proposed ERL-based fully coherent light source \cite{zhao2021energy}. It consists of a high-brightness photoinjector to generate a high-quality electron beam, followed by a main linac that provides acceleration to the target energy and enables energy recovery after radiation generation. Two 180° recirculating loops are used for beam transport and compression. Downstream of the first arc, a dedicated radiation section produces high-repetition-rate, ultrashort XFEL pulses. 
The quality of the electron bunches is determined by the injector, which is essential for overall operation. The electron beam is generated by a 550~kV DC gun 
\cite{nishimori2019operational} with a bunch charge of 20~pC. The cryogenic accelerating module consists of a string of eight 2-cell superconducting RF (SRF) cavities, which boosts the beam energy to 15 MeV. Using the tracking code $ASTRA$ \cite{floettmann2017space} in combination with a genetic algorithm, the injector is optimized to deliver a bunch with a normalized emittance of 0.2~mm-mrad and a peak current of 5.3~A at the injector exit, operating at a repetition rate of 1.3~GHz. The corresponding longitudinal phase space and current distribution are displayed in Fig \ref{fig:start}. The beam is transported through the merger system to the main linac, with the beam quality preserved effectively. The main linac comprises seven cryomodules, each consisting of a string of eight 9-cell SRF cavities. The first six cryomodules operate at 1.3~GHz and provide the main acceleration to the beam, with an average accelerating gradient of approximately 16~MV/m, consistent with the capabilities of state-of-the-art SRF systems \cite{liepe2012progress, sawamura2017superconducting}. A third-harmonic (3.9~GHz) cryomodule is employed to compensate for the nonlinear curvature induced by the main RF fields. As shown in Fig \ref{fig:acc}, the beam is accelerated to 1~GeV at the linac exit, where the longitudinal phase space remains approximately linear and provides favorable initial conditions for subsequent bunch compression. Downstream of the main linac, the beam enters the recirculating loop, which comprises two half-arcs. Each half-arc consists of two triple-bend achromat (TBA) cells, with all dipoles set to be 30°. As shown in Fig \ref{fig:arc}, the beam is compressed to about 240~fs (FWHM) after the first arc with a peak current of 85~A. To compensate for the projected energy chirp induced during large-angle bending and compression, an additional 3.9~GHz cryomodule is installed downstream of the arc to flatten the longitudinal phase space, as shown in Fig \ref{fig:flat}. Finally, the electron beam is transported into the radiation section to generate high-intensity ultrashort pulses. Three-dimensional S2E simulations have been carried out using $ELEGANT$ \cite{borland2000elegant} and $MAD$ \cite{grote1996mad} to demonstrate the acceleration and compression processes, taking into account the effects of the coherent synchrotron radiation (CSR) and the longitudinal wakefields. Key parameters in the injector, linac, and recirculating loop are listed in Table \ref{tab:1}.

\begin{table}
    \caption{Beam parameters in the injector, linac, and recirculating loop.}    
    \centering
    \begin{tabular}{ccc}
            \hline
            Parameter & Value & Unit \\  
            \hline
            Beam energy (injector) & 15 & MeV \\ 
            Beam energy (linac) & 1000 & MeV \\ 
            Normalized emittance & 0.2 & mm-mrad \\
            Bunch charge & 20 & pC \\
            Pulse duration (injector, FWHM) & 3 & ps \\
            Pulse duration (undulator, FWHM) & 240 & fs \\
            Peak current & 85 & A \\ 
            Relative energy spread & 0.015 & \% \\
            Repetition rate & 1.3 & GHz \\
            Bend angle in the ring & 30 & $^\circ$ \\          
            \hline
    \end{tabular}
    \label{tab:1}
\end{table}

\begin{figure}
\centering
\includegraphics[width=1\linewidth]{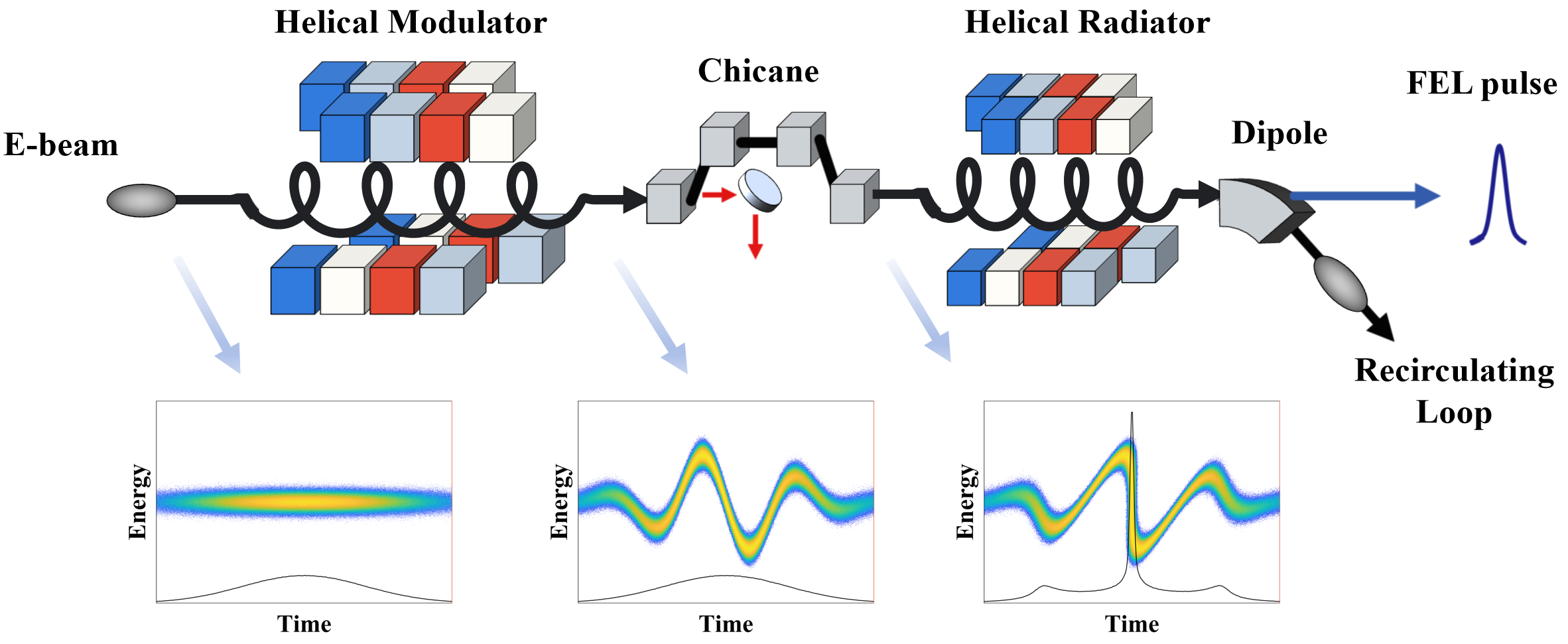}
\caption{\label{fig:layout}Schematic layout of the Radiation section.}
\end{figure}

The proposed scheme configuration (radiation section in Fig \ref{fig:erlall}), illustrated in Fig \ref{fig:layout}, comprises a long modulator, a large-dispersion chicane, and a short radiator. Rather than introducing an external optical laser, the upstream long helical modulator drives a continuous, phase-stable self-modulation process \cite{macarthur2019phase}. By setting the resonance wavelength in the THz regime, the interaction is effectively confined to the high-current core of the bunch. A helical wiggler is employed to provide symmetric transverse dynamics, mitigate the strong single-plane focusing, and increase transverse excursions that can arise at long resonance wavelengths. Crucially, because the resonance is in the THz regime, the current distribution of the electron beam is highly non-uniform within the scale of a single radiation wavelength. This inherent non-uniformity serves as an effective initial pre-bunching. Therefore, at the entrance of the modulator, rather than originating from random shot noise, the modulating field is directly driven by the beam's own coherently emitted THz radiation. Since the initial phase of this coherent radiation is strictly determined by the current profile, the ensuing self-modulation process is intrinsically phase-stable. This coherent radiation field, initially weak, is gradually amplified as it co-propagates with the electron beam. Within the extended modulation section, continuous energy exchange between the radiation and the electrons produces a robust, few-cycle energy modulation. Remarkably, the extended length of the wiggler enables the beam to naturally accumulate this strong modulation, even when starting with the intrinsically low peak current typical of ERLs. Since the interaction is strongest in the high-current core of the bunch, the modulation is naturally concentrated near the bunch center, while the head and tail remain only weakly modulated. After passing through a downstream dispersive section, this localized energy modulation is efficiently converted into an isolated high-current spike. Finally, the radiator section is further optimized to utilize this sharp current spike to achieve sufficient radiation power while suppressing satellite pulses.

\section*{Results and Discussion}

The feasibility of the proposed scheme was verified using the three-dimensional tracking code GENESIS 1.3 \cite{reiche1999genesis}. As illustrated in Fig \ref{fig:flat}, the electron beam develops a banana-like longitudinal phase-space distribution at the entrance of the modulation section due to nonlinear energy chirp and higher-order components in the compression process. Importantly, the longitudinal current profile remains approximately Gaussian, which is favorable for the proposed self-modulation mechanism. The modulator is composed of a 4~m-long in-vacuum helical undulator section with a period length of 14.5~cm. The $K$ parameter of the wiggler is 39.8 with a peak magnetic field of 2.9~T and a gap of 4.5~mm, leading to a resonant wavelength of 30~$\mu$m. Crucially, the electron bunch duration of approximately 240~fs is closely comparable to this modulator resonant wavelength (which corresponds to a temporal period of 100~fs). Because of this comparability, the electron beam current distribution is highly non-uniform within a single resonant wavelength. This macroscopic density gradient acts as an effective initial bunching, which inherently triggers the continuous, phase-stable self-modulation process.

\begin{figure}
\centering
\begin{subfigure}{\textwidth}
  \centering
  \includegraphics[width=0.75\linewidth]{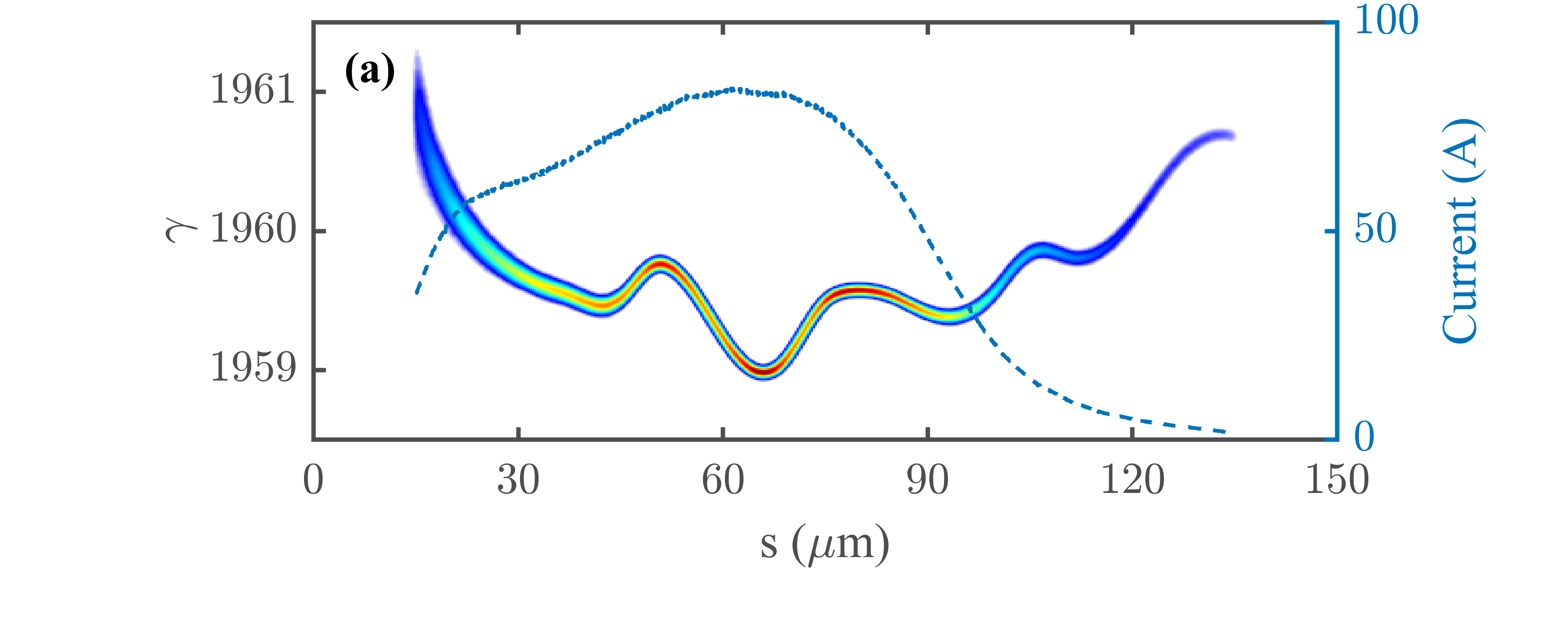}
  \label{fig:aftermodu}
\end{subfigure}

\begin{subfigure}{\textwidth}
  \centering
  \includegraphics[width=0.75\linewidth]{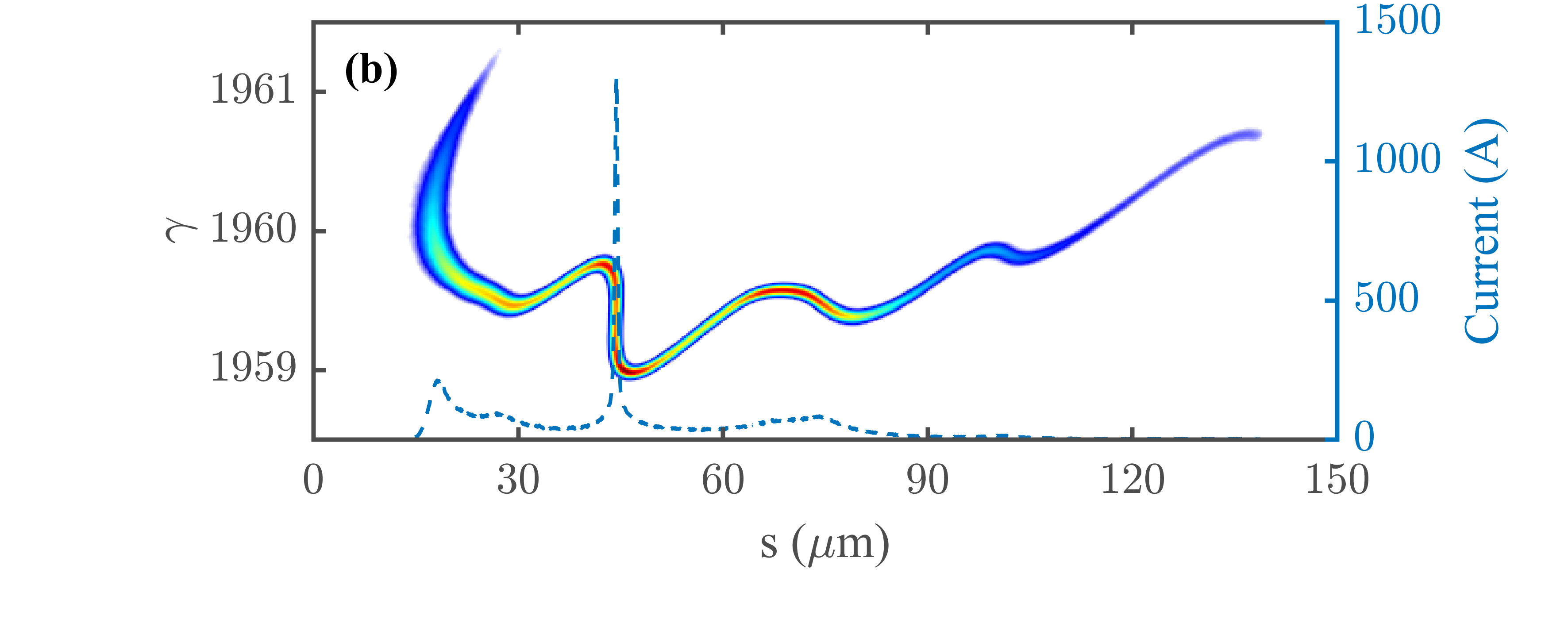}
  \label{fig:afterchicane}
\end{subfigure}

\caption{Longitudinal phase space and current distribution of the electron beam after (a) the modulator and (b) the chicane.}
\label{fig:modu}
\end{figure}

The longitudinal phase space and current distribution of the electron beam at the exit of the modulator are shown in Fig \ref{fig:modu}. The simulation results indicate that the THz radiation is progressively amplified along the modulator, reaching a power of 9.8~kW starting from the initial coherent emission driven by the macroscopic density gradient, rather than random shot noise. In the core region, the energy modulation induced by the interaction with the radiation field reaches 200~keV, which is about 16 times the initial slice energy spread. In the downstream dispersion section, an optimal $R_{56}$ value of 30~mm is required to convert the energy modulation into density modulation with a peak current of 1.3~kA and a pulse duration of about 1.9~fs (FWHM). Such a sharp high-current spike is essential for enabling the low-average-current ERL electron beam to produce high-peak-power FEL radiation.

\begin{figure}
\centering
\begin{subfigure}{.45\textwidth}
  \centering
  \includegraphics[width=\linewidth]{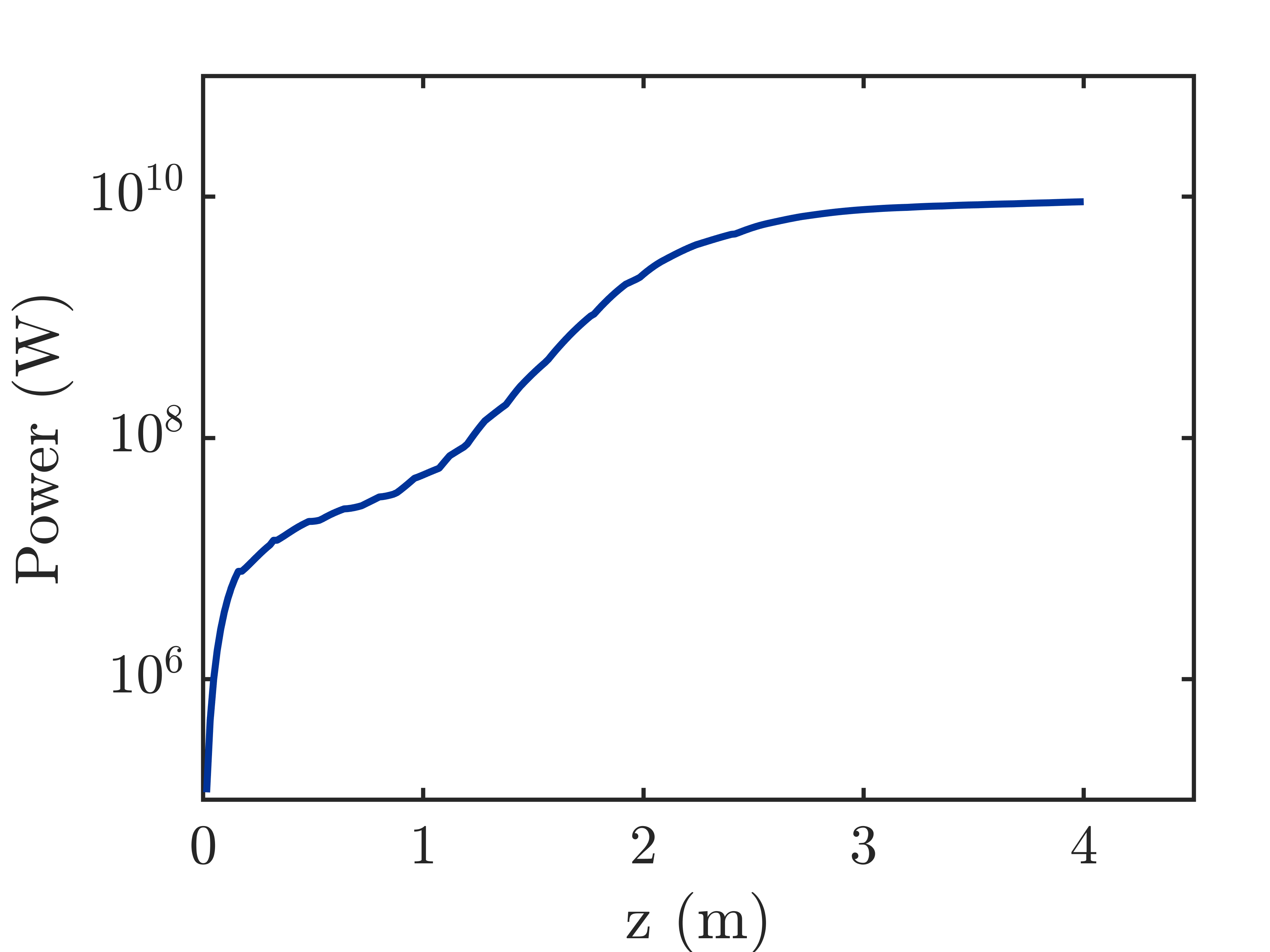}
  \caption{}
  \label{fig:gain}
\end{subfigure}%
\hfill
\begin{subfigure}{.45\textwidth}
  \centering
  \includegraphics[width=\linewidth]{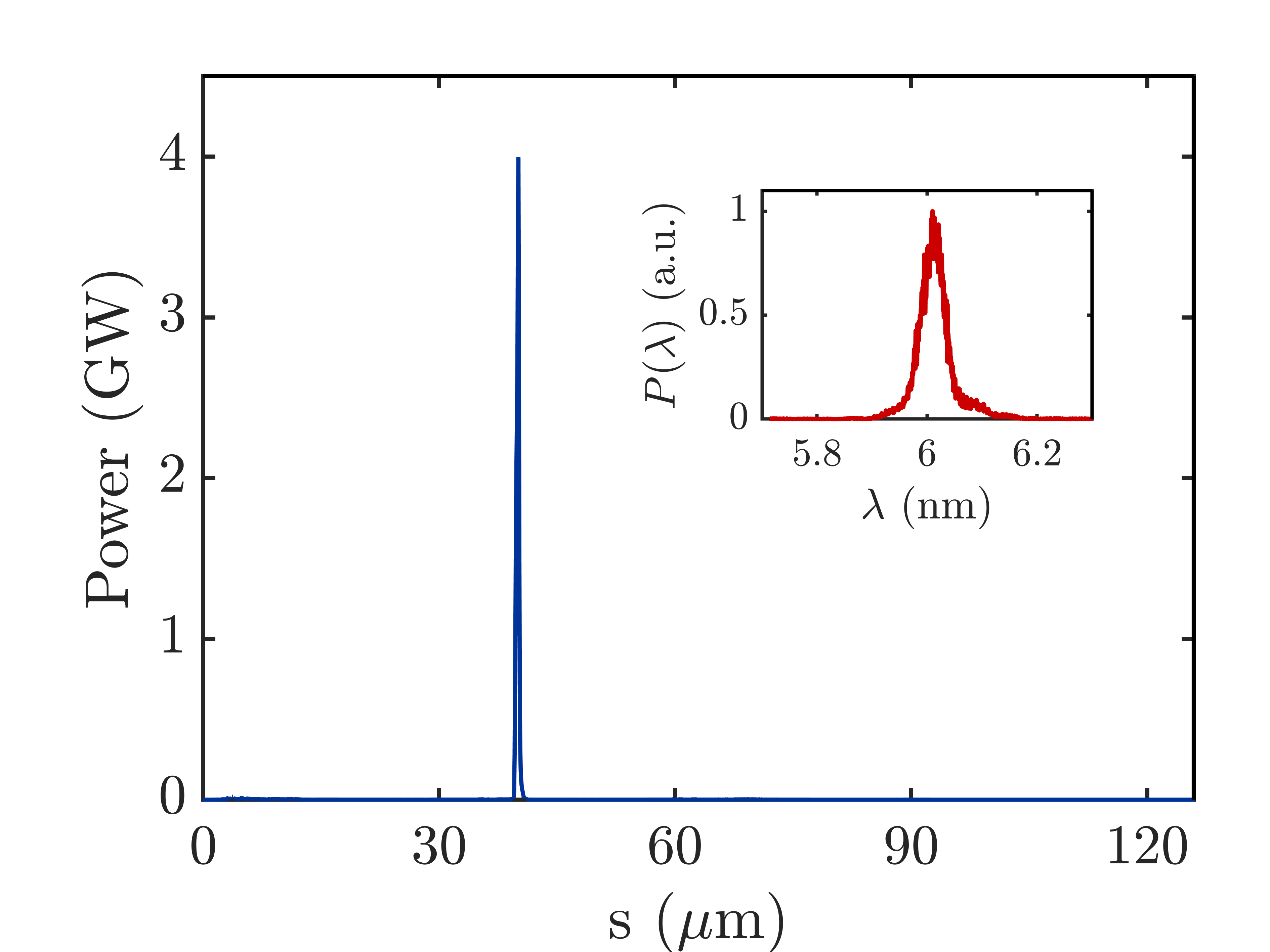}
  \caption{}
  \label{fig:rad_spec}
\end{subfigure}%
\caption{(a) FEL gain curve; (b) output radiation and spectrum at 2.24 m.}
\label{fig:rad}
\end{figure}

\begin{figure}[htbp]
  \centering

  \begin{subfigure}[b]{0.3\textwidth}
    \centering
    \includegraphics[width=\linewidth]{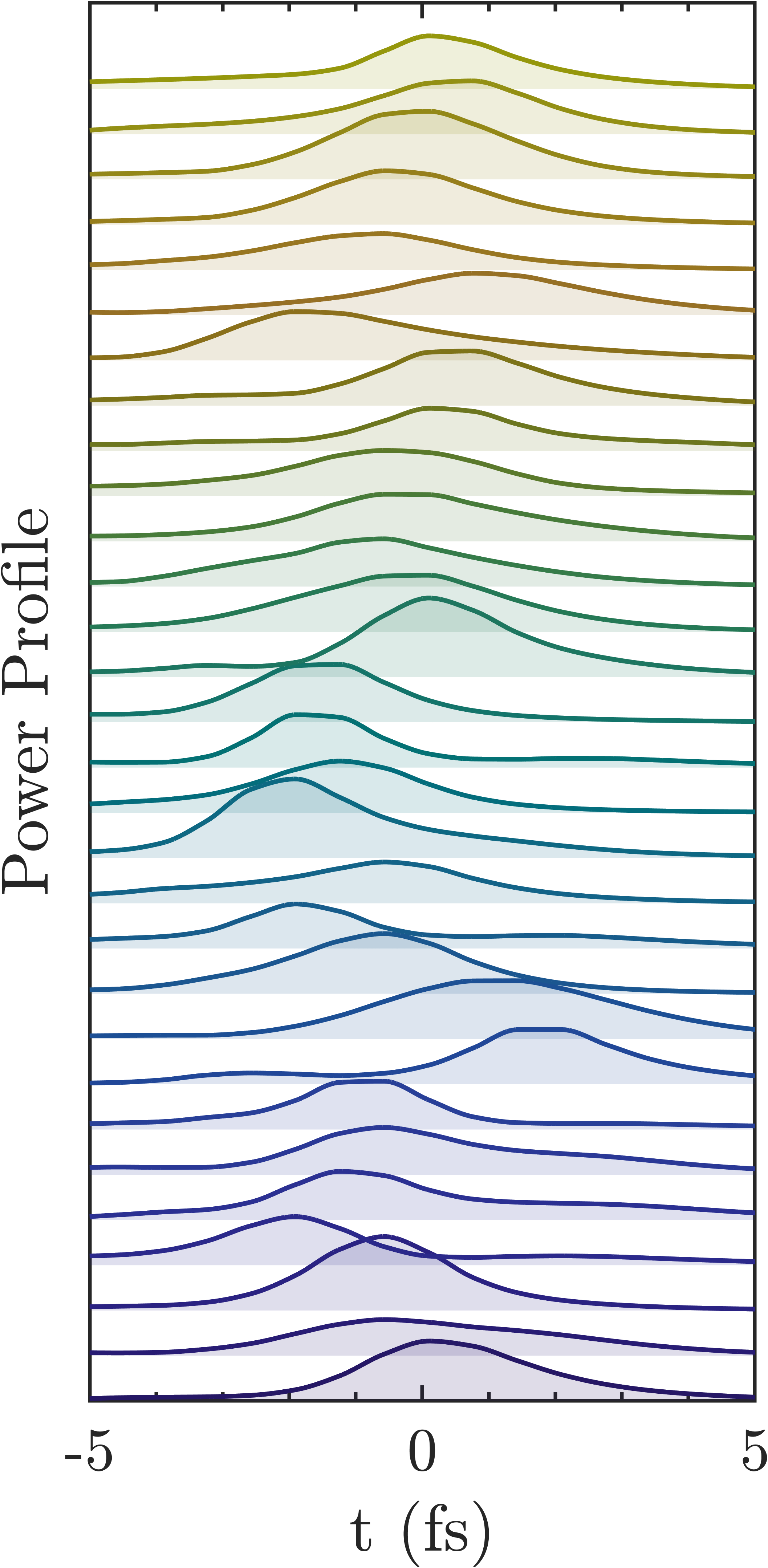}
    \caption{}
    \label{fig:powerfile}
  \end{subfigure}
  \hfill
  \begin{subfigure}[b]{0.3\textwidth}
    \centering
    \includegraphics[width=\linewidth]{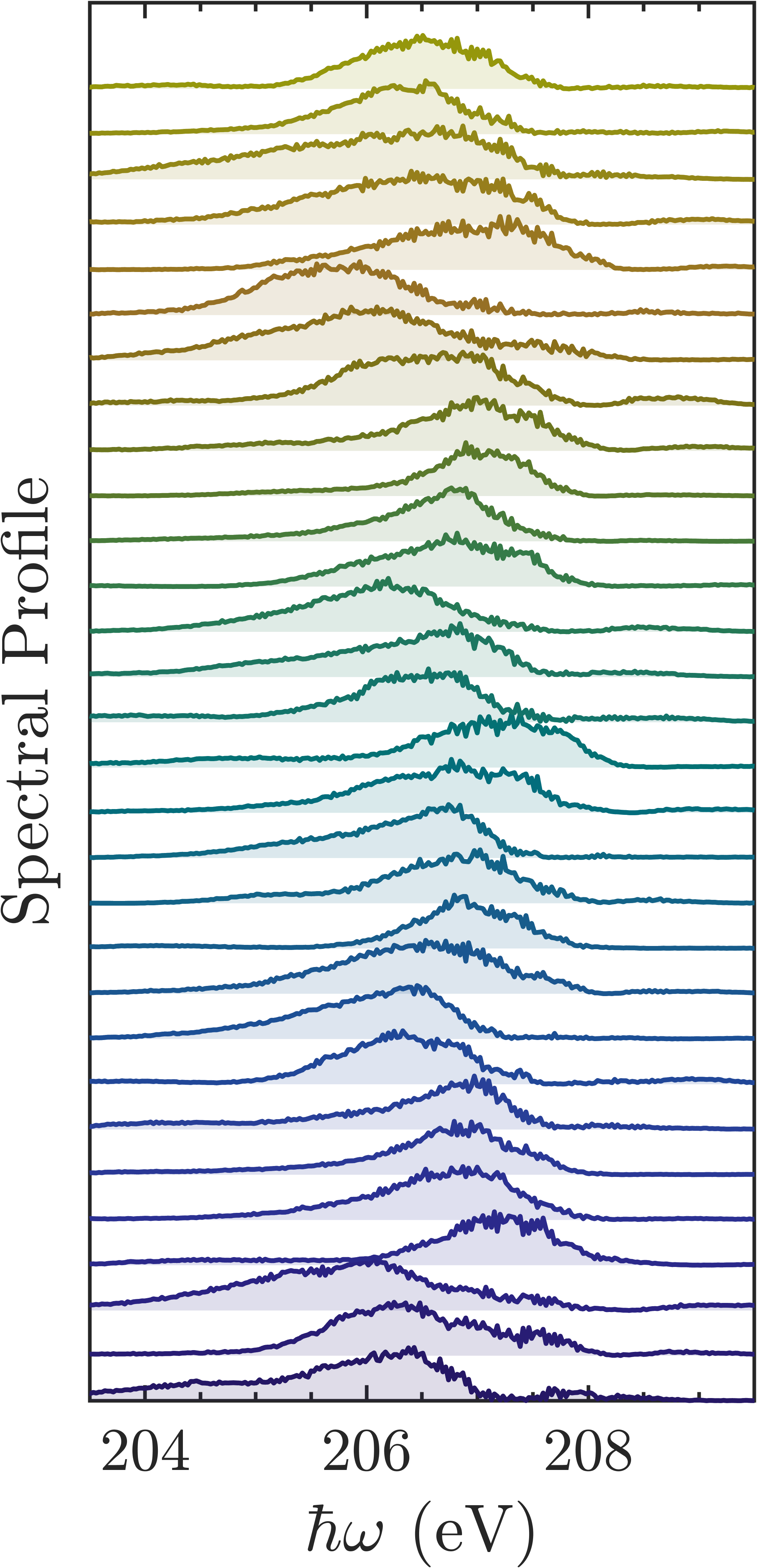}
    \caption{}
    \label{fig:specfile}
  \end{subfigure}
  \hfill
  \begin{subfigure}[b]{0.36\textwidth}
    \centering

    \begin{subfigure}{\textwidth}
      \centering
      \includegraphics[width=\linewidth]{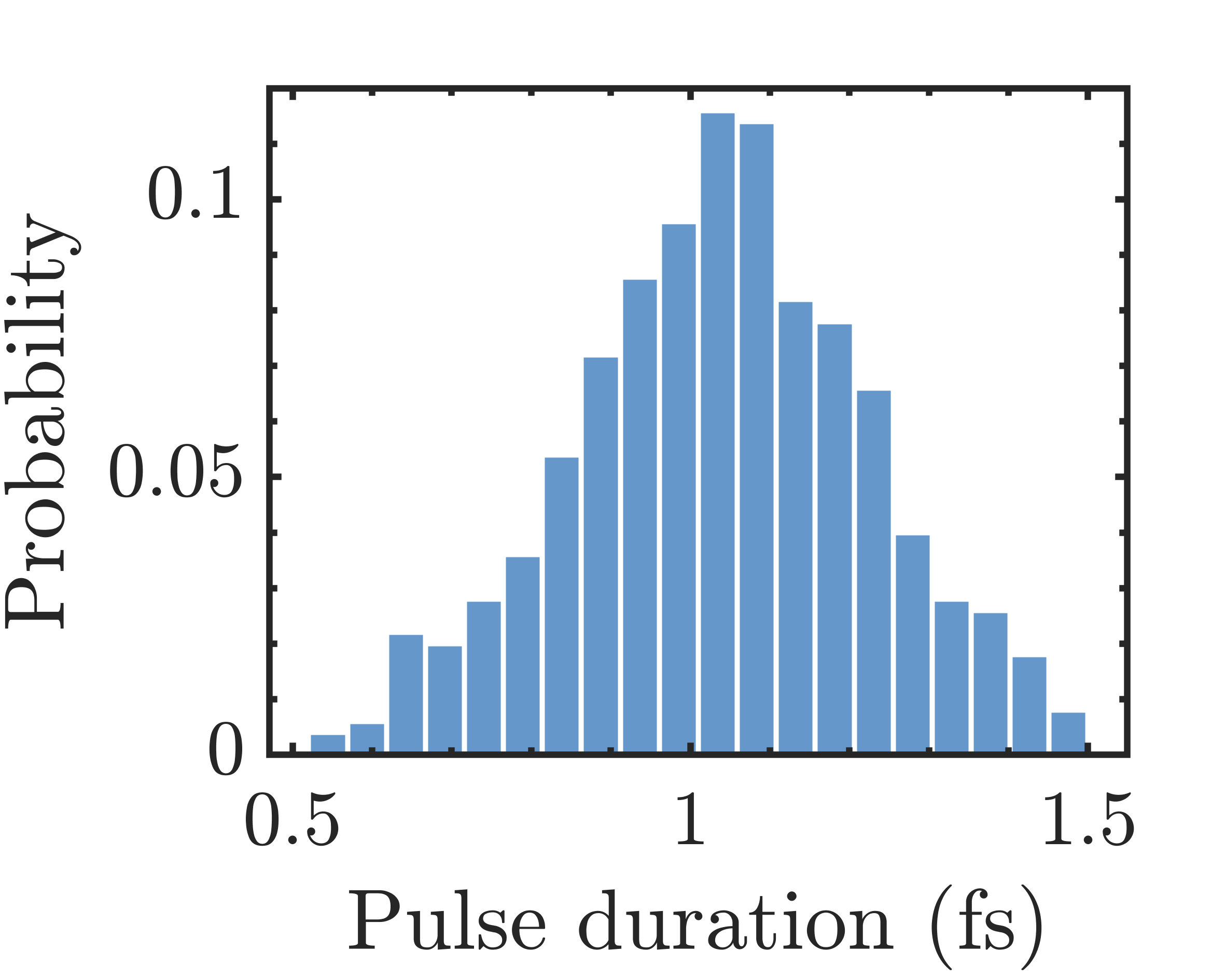}
      \caption{}
      \label{fig:powerfwhm}
    \end{subfigure}

    \vspace{1em} 

    \begin{subfigure}{\textwidth}
      \centering
      \includegraphics[width=\linewidth]{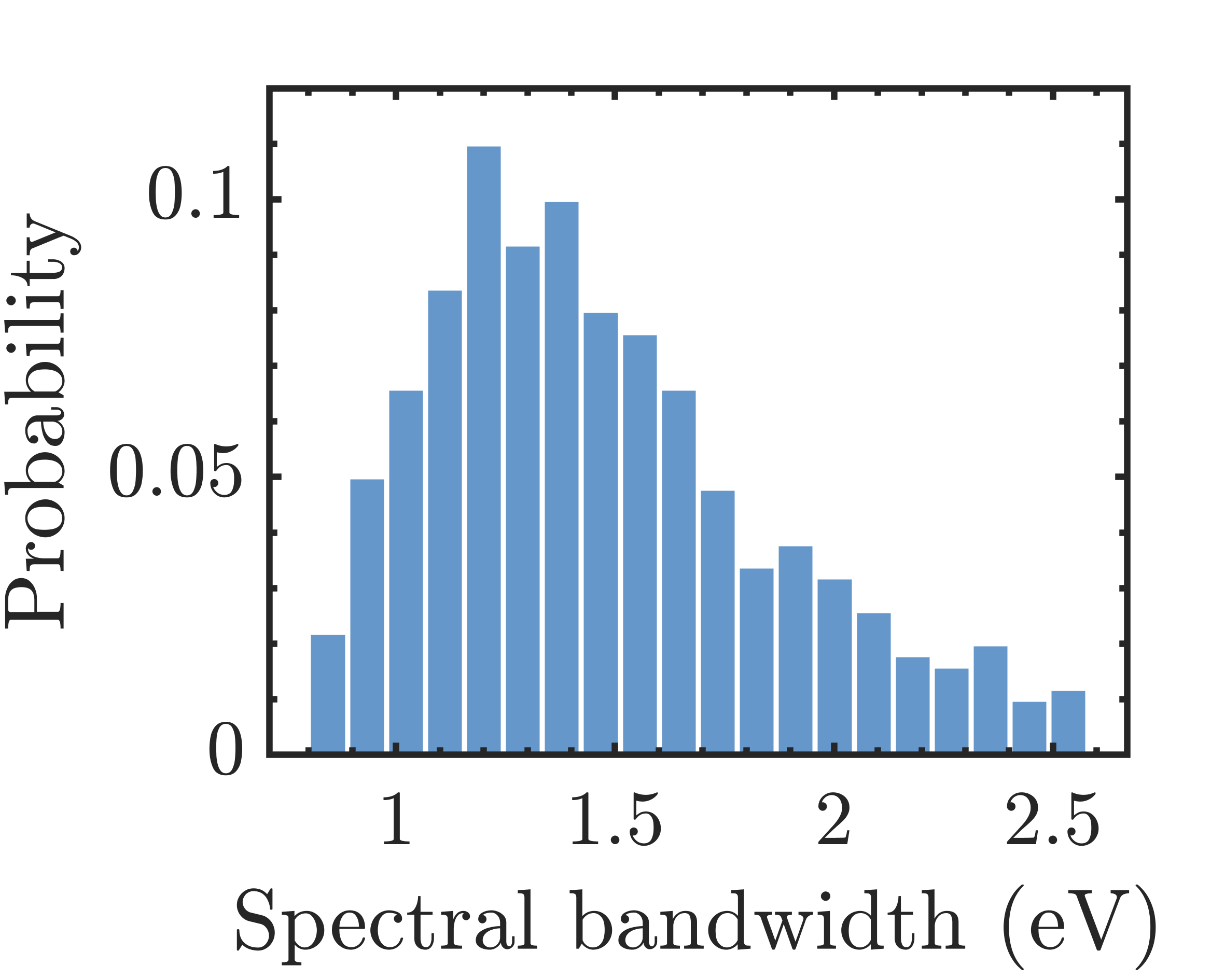}
      \caption{}
      \label{fig:specfwhm}
    \end{subfigure}

  \end{subfigure}

  \caption{Typical 30 shots of simulated (a) FEL power and (b) spectral profiles, with histograms of (c) the pulse duration (FWHM) and (d) the spectral bandwidth (FWHM) for 500 XFEL shots.}
  \label{fig:combined}
\end{figure}

\begin{figure}
\centering
\begin{subfigure}{.5\textwidth}  
  \centering
  \includegraphics[width=1\linewidth]{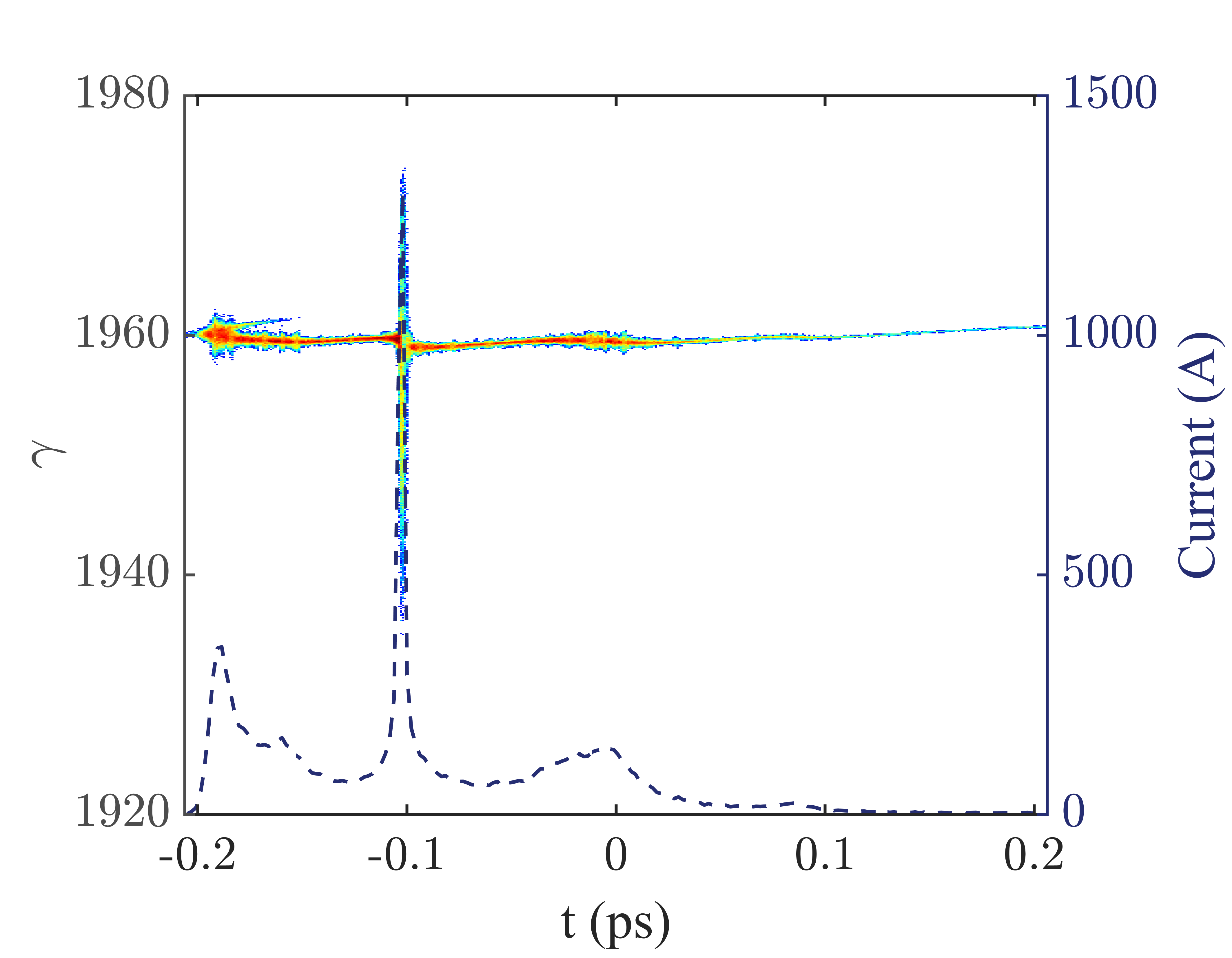}
  \caption{}  
  \label{fig:afterrad}
\end{subfigure}%
\hfill  
\begin{subfigure}{.5\textwidth}  
  \centering
  \includegraphics[width=1\linewidth]{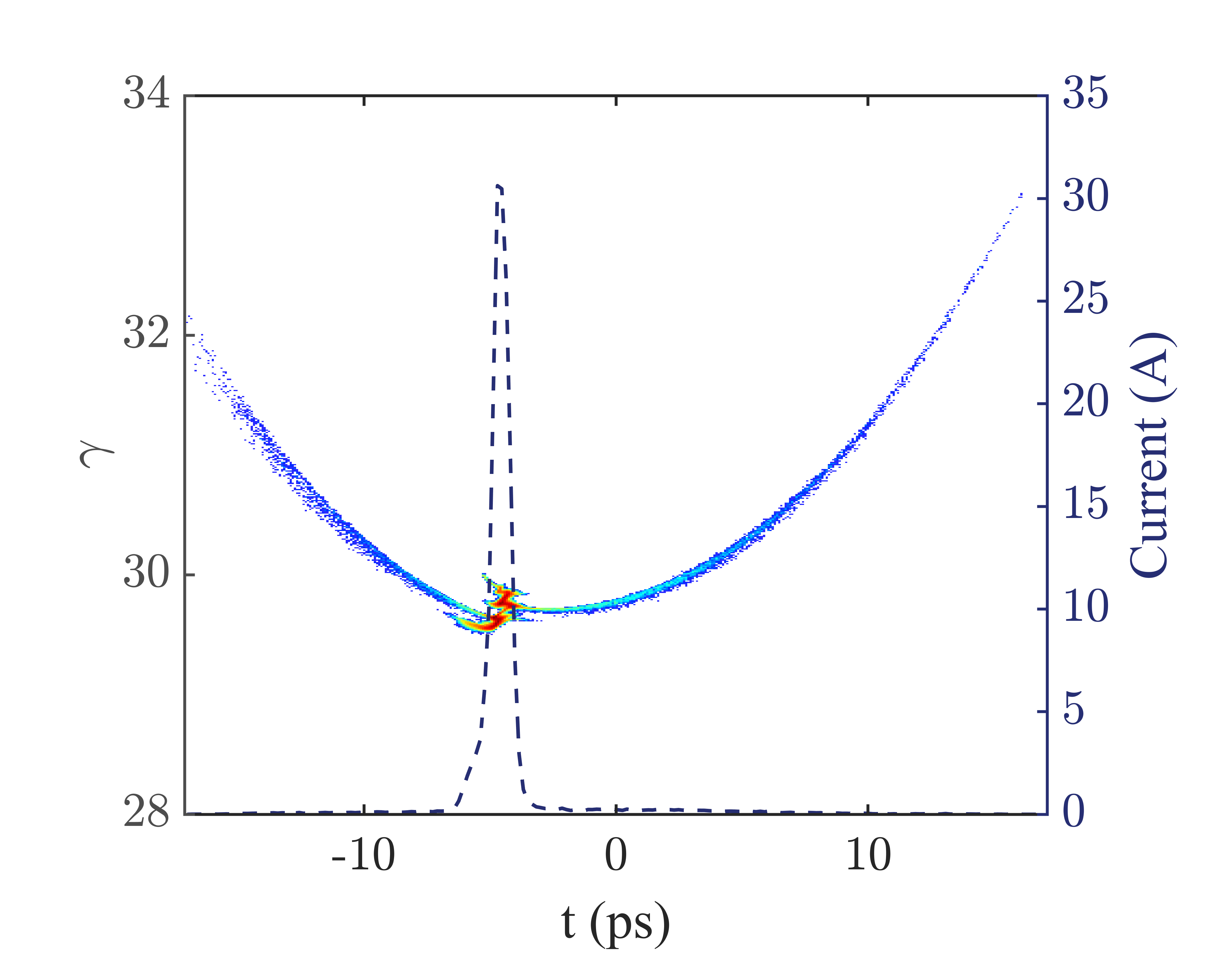}
  \caption{}  
  \label{fig:dump}
\end{subfigure}

\caption{Longitudinal phase space and current distribution evolution (a) after the radiator and (b) before the dump.}
\label{fig:recover}
\end{figure}
After the self-modulation accumulated in the modulator and dispersion section, the electron beam is transported into the radiator to generate high-intensity ultrashort radiation at the wavelength of 6 nm. The radiator comprises a helical undulator section with the period length of 16~mm. The undulator parameter K is~1.37, corresponding to a peak magnetic field of 0.91 T with a gap of 4.55 mm. As shown in Fig \ref{fig:gain}, the radiation power increases rapidly along the radiator and reaches saturation at approximately 3 m. The output radiation and spectral profiles are evaluated at z=2.24 m, where the main pulse has already reached the gigawatt level while satellite pulses and background radiation remain strongly suppressed, as shown in Fig \ref{fig:rad_spec}. At this optimized position, the radiation is generated with a peak power of 4 GW and a pulse duration of approximately 1.2~fs (FWHM). The spectral profile exhibits a FWHM bandwidth of 1.67~eV, corresponding to a relative bandwidth of 0.81\%. The time-bandwidth product is only 1.10 times the Gaussian Fourier-transform limit, confirming a nearly transform-limited pulse with high longitudinal coherence. In addition, the signal-to-noise ratio exceeds 99\%, confirming the strong dominance of the main pulse over the background radiation. The stability of the proposed scheme has been systematically investigated through a series of numerical simulations as well. By taking the initial shot noise into account, shot-to-shot fluctuations of the radiation performance have been analyzed. The distributions of 30 representative FEL pulses and their corresponding spectra are shown in Fig \ref{fig:powerfile} and \ref{fig:specfile}. Further statistical results from 500 shots are given in Fig \ref{fig:powerfwhm} and \ref{fig:specfwhm}. The simulation results indicate that the current spike forms at nearly the same longitudinal position within the electron bunch in each shot, leading to stable temporal localization of the emitted radiation pulse. The output pulses maintain gigawatt-level peak power and approximately 1 fs duration, with average values of 4.64 ± 1.04 GW for the peak power, 1.04 ± 0.19 fs for the pulse duration, and 1.47 ± 0.39 eV for the spectral bandwidth. In addition, the average pulse energy is 2.18~$\mu$J, corresponding to an average radiation power of 2.84~kW at the repetition rate of 1.3~GHz.

The energy recovery process is simulated using $ELEGANT$ from the exit of the radiator to the entrance of the dump, based on the six-dimensional electron distribution generated by $GENESIS~1.3$. The generated radiation is transported to the end-station for experimental use. The longitudinal phase space and current distribution of the electron beam along the downstream recirculating loop and during deceleration in the main linac are presented in Fig \ref{fig:recover}.  As shown in Fig \ref{fig:afterrad}, the lasing process introduces substantial radiation-induced energy spread and energy loss. As the electron beam passes through the second arc, the microbunching structure and the current spike formed in the radiation section are gradually smeared out, resulting in a peak current of approximately 30~A. Subsequently, the electron beam is returned to the main linac, where it is decelerated to 15~MeV with an average energy spread of 3\%. The bunch length is approximately 10~ps, slightly longer than that of the initial beam at the exit of the injector. The longitudinal phase space of the electron beam at the dump entrance is shown in Fig \ref{fig:dump}. According to the simulation, the energy recovery efficiency is about 99\%. Finally, the electron beam is delivered to the dump.

\section*{Conclusion}

In this study, we present a laser-free scheme for generating true continuous-wave, ultrafast XFEL pulses using an ERL-based light source. By exploiting a continuous, phase-stable self-modulation process driven by the beam's own coherent THz emission within an extended wiggler, the electron bunch naturally accumulates a robust energy modulation. Crucially, this modulation is subsequently converted into an exceptionally sharp, isolated current spike via a downstream dispersion chicane. This mechanism fundamentally overcomes the intrinsic low peak-current bottleneck of existing ERL technologies, providing a highly practical and immediate pathway to fulfill the stringent peak-current requirements for high-gain XFEL operation. 

Statistical S2E simulations demonstrate the highly stable generation of isolated soft X-ray pulses with peak powers exceeding 4~GW, pulse durations of about 1~fs (FWHM), and a remarkable average radiation power of 2.84~kW at an unprecedented 1.3~GHz repetition rate. Furthermore, with the ongoing development of multi-pass ERL technologies \cite{bartnik2020cbeta, angal2018perle}, the beam energy can be further increased without the need for additional acceleration modules. This advancement could make it possible to generate even shorter pulses and potentially extend the proposed scheme into the isolated attosecond regime. Ultimately, this scheme empowers current and future ERL facilities to deliver high-repetition-rate, sub-femtosecond X-ray pulses, offering a transformative platform for exploring ultrafast electronic and structural dynamics with broad applications in ultrafast spectroscopy, quantum materials research, and other frontier areas of X-ray science.
\section*{Acknowledgments}

The authors would like to thank Tao Liu,Yiwen Liu and Lu Cao for helpful discussions and useful comments.

\subsection*{Funding}
This work was supported by the National Natural Science Foundation of China (12435011), CAS Project for Young Scientists in Basic Research (YSBR-115), Strategic Priority Research Program of the Chinese Academy of Sciences (XDB0530000), and Shanghai Municipal Science and Technology Major Project.

\subsection*{Author Contributions} 
C.F. and Z.W. conceived the concept for the work. J.L. performed the simulations and analyzed the results with assistance from L.N. and Y.L. The manuscript was written by J.L. All authors reviewed the manuscript.

\subsection*{Competing interests}
The authors declare that they have no competing interests.

\subsection*{Data Availability}
The corresponding authors can provide relevant data if the demand is reasonable.

\printbibliography

\end{document}